# Modeling Soils' Failure Envelope with Virtual Displacements (LSVD)

Modélisation de l'enveloppe de rupture des sols avec déplacements virtuels (LSVD)

Emilio Fernando Altamirano-Muñiz
*National Autonomous University of Mexico (UNAM), Mexico City, Mexico*

2023-05-15

*2021360777.ipn@gmail.com

**ABSTRACT:** This work proposes the Least Squares with Virtual Displacements (LSVD) method to obtain the shear strength of *n* soil samples for the cohesionless, frictionless and mixed resistance conditions. Finding a common tangent line of more than two samples may not have an exact solution, especially if there are measurement errors. LSVD is an iterative method to find the best common tangent line or function to the *n* soil samples. There are soils that can be modeled with a linear failure envelope but the LSVD can be formulated in different forms, adapting itself to other criterion too, like the logarithmic failure envelope. The tested data is analyzed with LSVD and compared between other commonly used methods, like p-q method and CTPAC. The soils' samples were subjected to random noise to simulate errors from measurements to study the method's results. Also with that analysis a common reference point can be found to compare the methods given that all of them take different points of the Mohr Circles.

**RÉSUMÉ**: Ce travail propose la méthode des moindres carrés avec déplacements virtuels (LSVD) pour obtenir la résistance au cisaillement de n échantillons de sol pour les conditions de résistance, sans cohésion, sans frottement et mixte. Trouver une ligne tangente commune de plus de deux échantillons peut ne pas avoir de solution exacte, surtout s'il y a des erreurs de mesure. LSVD est une méthode itérative pour trouver la meilleure ligne ou fonction tangente commune aux *n* échantillons de sol. Il existe des sols qui peuvent être modélisés avec une enveloppe de rupture linéaire mais le LSVD peut être formulé sous différentes formes, s'adaptant également à d'autres critères, comme l'enveloppe de rupture logarithmique. Les données testées sont analysées avec LSVD et comparées entre d'autres méthodes couramment utilisées, comme la méthode p-q et le CTPAC. Les échantillons de sols ont été soumis à un bruit aléatoire pour simuler les erreurs de mesure afin d'étudier les résultats de la méthode. Cette analyse permet également de trouver un point de référence commun pour comparer les méthodes étant donné que toutes prennent des points différents des cercles de Mohr.

## 1 INTRODUCTION

One of the important procedures when analyzing soils is determining their resistance parameters, i.e., friction angle and cohesion through the failure envelope (Labuz and Zang, 2012).

Finding the failure envelope is not an straight task, and approximations are used. Even though exact solutions may exist, theoretically that should be the case, errors are carried in the laboratory measures and the continuum hypotheses are definitely not fulfilled, so the possibility of finding a line that is tangent to all tested samples becomes likely impossible, and even if there was, there is no guarantee it is the real failure envelope. Furthermore, a linear envelope is an approximation given that soils' strength have a non linear behaviour (Maksimovic, 1989).

To apply a linear regression a data set *D* is needed with labeled data, although after the laboratory analysis of a soil we get data such as $\sigma_1$ and $\sigma_3$, that is not a straight input to just apply a linear regression. That is why some different methods have been suggested, all of them (including this one) procure the tangent failure theory.

A good approximation is the *p-q* method due to its precision and the ease of calculation, which is a linear regression of the center of the circles $\lambda_i$ and the radii $r_i$ (Wu and Tung, 2020). It is a good approximation because the real failure point happens to be close to the center of the circle, therefore the difference is not quite large. But another characteristic is that it underestimates the soil's shear strength because the linear functions computed from those points will have smaller slopes than where the tangency should be, which is certainly better than to overestimate it, but surely is not accurate and does not take into account the Mohr-Coulomb failure theory (Mukherjee, 1987).

Other method that will be compared here is the CTPAC (Wang, 2004), which finds the common tangent between two adjacent circles and does that for all circles then applies a linear regression to all the calculated points. An advantage of that method is that if there is an exact solution (unlikely), it can find that solution with only one pair of circles.

Along this works, the notation will be from the statistical jargon. $\tau$ is the shearing stress, $\hat{\tau}$ is the estimation of the failure envelope, $\overline{()}$ is the arithmetic mean of $()$, $\beta_j$ are the estimation coefficients, $\lambda_i$ is the center of a circle, and $r_i$ is the radius of that same circle. The $i$ subindex is associated to each circle and goes to $n$ number of samples.

## 2 FORMULATION

In order to apply least squares, the loss function ought to be established as it is shown in equation (2.1).

$$\ell(\tau, \hat{\tau}) = \sum_i (\tau_i - \hat{\tau}_i)^2 \qquad (2.1)$$

The $\hat{\tau}_i$ is the assumed shearing failure envelope model, but $\tau_i$ it is unknown too, in statistical learning problems $\tau_i$ comes from a labeled data set and should be points where the envelope touches the circles, the circles' functions can be obtained but the failure points are yet to be known, so for this method is proposed to apply *virtual displacements*. This idea comes from formulating the problem of finding a linear function that is tangent to all circles *or* minimizes the error between the *virtual tangent*, i.e, the loss function will compare the point $^\theta\sigma_i$ from $\hat{\tau} = \beta_1{}^\theta\sigma_i + \beta_0$ with the semi-circle function $\tau_i = \left(r_i^2 - ({}^\theta\sigma_i - \lambda_i)^2\right)^{0.5}$. But to know which point to take as $^\theta\sigma_i$ the equation (2.2) is solved, in other words, finding where *would* it be tangent:

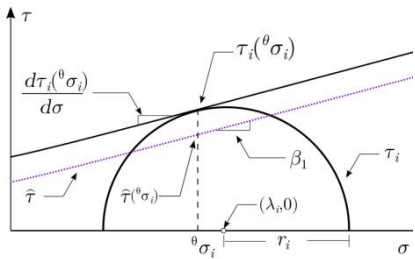

*Figure 1. Scheme of the virtual displacements by the $^\theta\sigma_i$ to compute the square error.*

$$\frac{d\tau_i}{d\sigma} = \frac{d\hat{\tau}_i}{d\sigma} \qquad (2.2)$$

### 2.1 Linear Failure Envelope

This is the case of the Mohr-Coulomb linear criteria, so the estimation $\hat{\tau} = \beta_1 \sigma + \beta_0$ is the failure envelope, where the bias $\beta_0$ and $\beta_1$ happens to be the cohesion and friction angle's tangent, respectively. Then $\hat{\tau}'$ is $\beta_1$ and the point where the virtual displacement will happen will satisfy equation (2.3).

$$\frac{d\tau_i}{d\sigma} = \beta_1 \qquad (2.3)$$

The solution to equation (2.3) is the equation (2.4).

$$^\theta\sigma_i = \lambda_i - \frac{\beta_1 r_i}{\sqrt{\beta_1^2 + 1}} \qquad (2.4)$$

It is only taken the solution that makes sense with the failure envelopes. Thus the loss function is

$$\ell(\tau, \hat{\tau}) = \sum_i \left\{ \sqrt{r_i^2 - ({}^\theta\sigma_i - \lambda_i)^2} - \beta_1\,{}^\theta\sigma_i - \beta_0 \right\}^2 \qquad (2.5)$$

So by equating the gradient with the zero vector $\nabla \ell(\tau, \hat{\tau}) = \overline{0}$ the systems of equations can be solved. Then, from the systems of linear equations, it can be easily obtained $\beta_0$ but not that straight $\beta_1$:

$$\beta_0 = \bar{r}\sqrt{\beta_1^2 + 1} - \beta_1 \bar{\lambda} \qquad (2.6)$$

So the equation to find $\beta_1$, and effectively find the envelope, is the equation (2.7):

$$0 = \beta_1 \sum_i (\lambda_i^2 + r_i^2) + n\beta_0 \left(\bar{\lambda} - \bar{r}\beta_1/\sqrt{\beta_1^2 + 1}\right) - \left\{(2\beta_1^2 + 1)/\sqrt{\beta_1^2 + 1}\right\} \sum_i (\lambda_i r_i) \qquad (2.7)$$

When $n = 1$, given any $\beta_1$, equation (2.7) will always sum zero, so it is only a matter of calculating the $\beta_0$ or $\beta_0$ can be fixed and just compute the $\beta_1$. But one sample goes beyond practical applications.

### 2.2 Logarithmic Failure Envelope

The logarithmic equation has the form $\hat{\tau} = \beta_1 \ln(\sigma + \beta_2) + \beta_0$, therefore, its virtual displacement is given by equation (2.8), which has solutions by radicals but it is a very long solution. Noted this, a good method to solve it is the bisection method, also taking into account where the solution must lie.

A note for the virtual displacement solution of (2.2) is that, depending of the implemented method like the logarithmic criteria, equation (2.2) may have various solutions, some may even be in the complex plane. Then, the solution that we only care for is the one that must lie within the ($^i\sigma_3, \lambda_i$] interval, if the solution is $\lambda_i$ then it is the frictionless condition.

$$0 = \frac{-\sigma + \lambda_i}{\sqrt{r_i^2 - (\sigma - \lambda_i)^2}} - \frac{\beta_1}{\sigma + \beta_2} \quad (2.8)$$

And the loss function $\ell(\tau, \hat{\tau})$ can be minimized and its coefficients computed, by doing so, the problem is solved:

$$\overline{\beta} = \arg\min_{\beta_0, \beta_1, \beta_2} \ell(\tau, \hat{\tau}) = \sum_i \left\{ \sqrt{r_i^2 - (^\theta\sigma_i - \lambda_i)^2} - \beta_1 \ln(^\theta\sigma_i + \beta_2) - \beta_0 \right\}^2 \quad (2.9)$$

## 3 RESULTS

The samples' data is shown in table 1 and the tests were conducted on intact samples, sample's 1 data was obtained through unconsolidated undrained triaxial strength tests on clay, and sample 2 data (Guo et al., 2020) is from the conventional triaxial tests on rocks. The LSVD's results shown on figure 2 were computed from sample 1 by solving equation (2.7) and the LSVD's solution of figure 3 was computed by optimizing equation (2.9), noting that in each iteration of the optimization algorithm equation (2.2) is solved for each circle. Given that an exact solution is unlikely and is no guarantee of being the real failure envelope, and each methods compares from different points, then it is suitable to analyze their variability and confidence intervals as shown in figure 4 and 5.

*Table 1. Failure pressures of samples [MPa]. * Test data from Guo et al., 2020*

| Sample | | | | | |
|---|---|---|---|---|---|
| 1 | $\sigma_3$ | 0.04903 | 0.09806 | 0.14709 | 0.19613 |
| | $\sigma_1$ | 0.21686 | 0.31795 | 0.38337 | 0.47907 |
| 2* | $\sigma_3$ | 5 | 20 | 40 | 80 |
| | $\sigma_1$ | 100 | 154 | 221 | 323 |
| 1 | $\sigma_3$ | | | | |
| | $\sigma_1$ | | | | |
| 2* | $\sigma_3$ | 90 | 100 | | |
| | $\sigma_1$ | 346 | 361 | | |

Laboratory measurements will carry errors due to the equipment, environment or other variables, so the samples' data were put under random noise of 10%, $\epsilon \sim$ Unif [-0.1, 0.1]. The random noise is also used to analyze the response of the LSVD with sample errors, without standing the already existing ones from the readings, i.e, it is going to be assumed that the measures have zero error initially and they are the true parameters.

Figure 2 and 3 show the samples with their failure envelopes corresponding to their methods. Figure 4 are the confidence intervals with two standard deviations of the linear *p-q*, linear LSVD and the linear CTPAC. Figure 5 are the confidence intervals with two standard deviations of the logarithmic *p-q*, logarithmic LSVD, and logarithmic CTPAC.

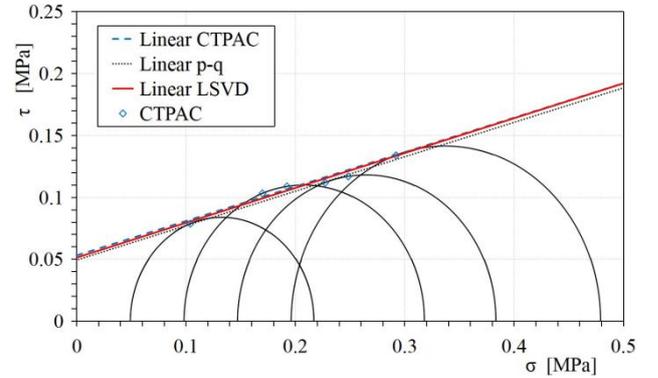

*Figure 2. Failure envelope computed with LSVD, CTPAC and p-q of sample 1.*

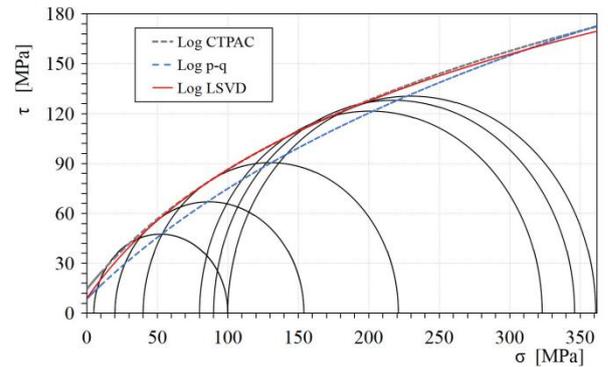

*Figure 3. Failure envelopes computed with logarithmic LSVD, CTPAC and p-q of sample 2.*

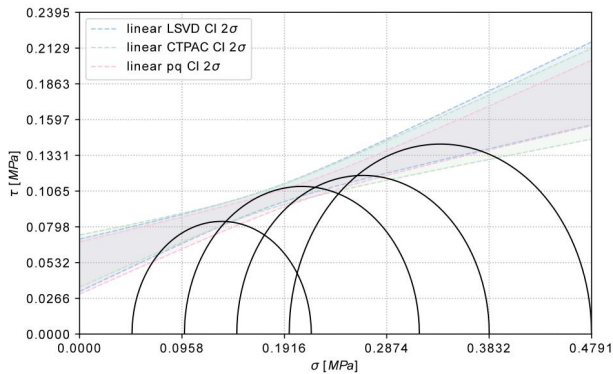

*Figure 4. Sample 2: Confidence intervals of the linear LSVD, p-q and CTPAC, considering a 10% of random noise of sample 1.*

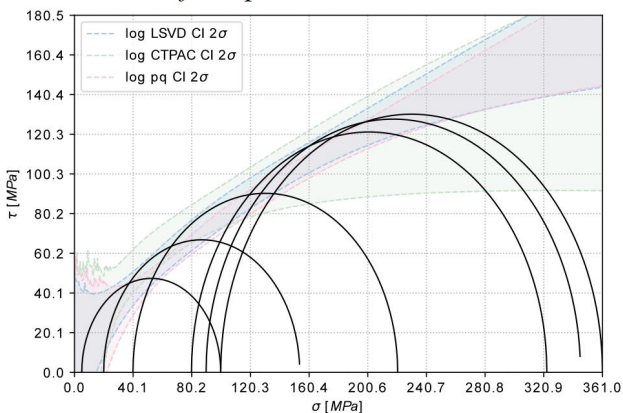

*Figure 5. Sample 2: Confidence intervals of the logarithmic LSVD, p-q and CTPAC, considering a 10% of random noise of sample 2.*

## 4 DISCUSSION

The LSVD method takes into account the tangency within its process even if the result is a secant line or does not touches a circle, while *p-q* ignores it as a trade-off of simplicity. CTPAC relies on the fact that a tangency is shared between all circles, which is the core theoretical assumption and gives good results, especially on the linear case.

Adding random noise to the samples help to simulate measurement errors and in such way visualize the resistance variations of the soils. More samples would give finer results, three to six samples are proper for practical applications.

Although, visually we can asses the good fit of each method, it becomes hard to take a comparison point due to the nonexistence of exact solutions. If the methods consist on regression techniques, therefore all of them will have a good fit relative to their reference points, e.g. virtual tangents (LSVD), adjacent tangent points (CTPAC) or radii (*p-q*). Having this in mind, the confidence interval is useful to check how sensitive the models are. So if the confidence interval at two standard deviations captures the sections where the tangency surely must happen then it is better than not capturing it. The *p-q* does not capture where the tangent line must be on the linear and logarithmic test with the samples, CTPAC surely captures the sections but has a great variability which is the opposite of precision and is undesired, LSVD behaved more precise after the random noise.

## 5 CONCLUSIONS

The LSVD obtains the failure envelope considering the Mohr-Coulomb failure theory and withdraws a lot of uncertainty coming from the mere methods that are applied, remaining the inconveniences of assuming the elasticity hypotheses as true for a soil.

The linear LSVD is easy to implement in a spreadsheet or coding it, so it is feasible to use. Also LSVD is able to find the exact solution (if it exists), but if there is one, then CTPAC is a lot quicker for the linear case due to the need of only one calculation. Mentioning the existence or not of exact solutions helps to the problem formulation but it should not be important because even if an exact solution is found (a perfectly tangent function), that does not mean the failure envelope is the real one, recalling the presence of errors.

LSVD can be harder to implement numerically to other nonlinear failure envelope criterion due to solution of equation (2.2) because it might have various solutions and they change every iteration, it can be stated that it uses *dynamic* points, while with CTPAC and *p-q* have *static* failure points and make it easier to apply any estimation function.

REFERENCES

Guo, B., L. Wang, Y. Li, and Chen, Y. (2020) Triaxial strength criteria in mohr stress space for intact rocks, *Advances in Civil Engineering*, vol. 2020, pp. 1–13. https://doi.org/10.1155/2020/8858363

Labuz F., Zang, A. (2012) Mohr–Coulomb Failure Criterion. *Transportation Infrastructure Geotechnology*, vol. 7, 03. https://doi.org/10.1007/s00603-012-0281-7

Maksimovic, M. (1989) Nonlinear Failure Envelope for Soils, *Journal of Geotechnical Engineering*, vol. 115, pp. 581–586. https://doi.org/10.1061/(ASCE)0733-9410(1989)115:4(581)

Mukherjee, A. (1987) *Methods For Determining Shear Strength Of Soils And The Limitations And/Or Advantages Of The Various Tests*, Master thesis, Kansas State University. [online] Available at: [https://core.ac.uk/reader/33353980]

Wang, X.H. (2004) Regression solution for general stress intensity of the tri-axis shearing. *Journal of Earth Science and Environment,* vol. 2004-1, pp. 52–54.

Wu, J., Tung, C.Y. (2020) Determination of model parameters for the hardening soil model, *Transportation Infrastructure Geotechnology*, vol. 7, 03. https://doi.org/10.1007/s40515-019-00085-8

## A  APPENDIX

The script to use LSVD can be found on the following github repository, it is written in Python. The required libraries are `numpy` and `scipy`, although `matplotlib` is also used to plot the results.